\documentclass[useAMS,usenatbib]{mn2e}
\usepackage{psfig}
\def\ltsima{$\; \buildrel < \over \sim \;$}
\def\simlt{\lower.5ex\hbox{\ltsima}}
\def\gtsima{$\; \buildrel > \over \sim \;$}
\def\simgt{\lower.5ex\hbox{\gtsima}}
\def\cgs{{erg cm$^{-2}$ s$^{-1}$}}
\def\ergs{{erg s$^{-1}$}}
\def\cm2{{cm$^{-2}$}}
\def\xnu{{$\chi^{2}(\nu)$}}

\def\fhx{{$F_{2-10}$}}
\def\lum{{$L_{2-10}$}}
\def\p1{{Paper I}}
\def\fsx{{$F_{0.5-2}$}}
\def\xmm{{\em XMM--Newton}}
\def\chandra{{\em Chandra}}
\def\nhgal{{N$_{\rm H}^{\rm Gal}$}}
\def\nh{{N$_{\rm H}$}}
\def\epic{{\em EPIC}}

\def\pn{{\em PN}}
\def\mos{{\em MOS}}
\def\nhc{{N$^{T4}_{\rm H}$}}  
\def\nhh{{N$^{T5}_{\rm H}$}} 
\def\lums{{$L_{0.5-2}$}}
\title[Evidence for a multi--zone warm absorber in MKN 304]{Evidence for a multi--zone warm absorber in the XMM--Newton spectrum of Markarian 304}
\author[E. Piconcelli et al.]{E. Piconcelli$^{1}$\thanks{E-mail: epiconce@xmm.vilspa.esa.es}, E.~Jimenez-Bail\'{o}n$^{1}$, M.~Guainazzi$^{1}$, N.~Schartel$^{1}$, 
\newauthor P.M.~Rodr\'{i}guez-Pascual$^{1}$, M.~Santos-Lle\'{o}$^{1}$\\
$^{1}$ XMM-Newton  Science Operation Center/RSSD--ESA, Apartado 50727, Madrid 28080, Spain\\}

\begin{document}

\date{Accepted. Received}

\pagerange{\pageref{firstpage}--\pageref{lastpage}} \pubyear{2003}

\maketitle

\label{firstpage}

\begin{abstract}
We present a \xmm~observation of Markarian 304, a Seyfert 1
 galaxy at $z$ = 0.066. The \epic~data
 show that MKN~304 is affected by heavy (\nh~$\approx$ 10$^{23}$ \cm2)
 obscuration due to ionized gas.   A two--phase warm absorber provides
 an adequate parameterization of this gas.  The  ionization
 parameter of the two components is $\xi \approx$ 6 \cgs~and $\xi
 \approx$ 90 \cgs, respectively. The observed continuum photon index
 ($\Gamma \approx$ 1.9) is typical  for Seyfert 1 galaxies. Two
 significant emission lines are detected at 0.57 keV and 6.4 keV,
 respectively.  The former is mostly likely due to He--like oxygen 
triplet emission arising from an ionized plasma (maybe the warm absorber itself).
  The latter is due to fluorescent emission of K--shell iron in a
 low--ionization state (FeI--XV). The upper limit for the line width
 of $\sigma_{K\alpha} <$ 0.18 keV most likely rules out an origin in
 the inner parts of the accretion disk. 
 Interestingly, the strength of such line is consistent with
 the possibility that the emission is produced in the warm absorber itself.
 However, a  substantial contribution from the torus is plausible too. We
 have also found a weak (4\% of the primary continuum) soft excess
 emission component.  The presence of this excess could be explained
 by either emission/scattering from a warm gas or partial covering, or
 a combination of them.
\end{abstract}

\begin{keywords}
galaxies: active -- galaxies: individual: Markarian 304 -- galaxies: Seyfert -- X--rays: galaxies.
\end{keywords}

\section{Introduction}
The bulk of the X--ray emission in active galactic nuclei (AGNs) is
produced via Comptonization of UV photons in the inner parts of the
accretion flow close to the central supermassive black hole.  The
resulting power law spectrum is modified by a number absorption and
emission spectral features due to the reprocessing of this primary
continuum (Mushotzky, Done \& Pounds 1993). They are clear signatures 
of the presence  of  large amounts of gas characterized by different
physical properties in the circumnuclear region.

One of the most common spectral features is the so--called
{\it warm} (i.e. partially ionized) {\it absorber}  (Halpern 1984; Pan
et al. 1990, Turner et al. 1993). Such a component has been found to
contribute significantly to the opacity of the X--ray primary emission
in $\approx$ 50\% of the Seyfert 1 galaxies with column densities up
to \simgt~10$^{23}$ \cm2 (Reynolds 1997).  In particular, OVII (0.739
keV) and OVIII (0.871 keV) absorption edges resulted to be the most
prominent  absorption signatures of the ionized gas in low--resolution
{\em ASCA} observations.

The advent of  grating spectrometers on--board \xmm~and \chandra~has
dramatically enriched our knowledge in this field. High--resolution
soft X--ray spectra of bright Type 1 AGNs present a wealth of
absorption lines (e.g. Collinge et al. 2001; Kaspi et al. 2001;
Kaastra et al. 2002),  which allow to infer a few remarkable
conclusions: ($i$) warm absorbers usually show a very complex
structure, with  multiple regions characterized by different
ionization states,  column densities and  velocities;  ($ii$) the
strongest features are due to He-- and H--like states of O, Ne, Mg and
Si; in addition a blend of M--shell iron  inner shell transition
(unresolved transition array or UTA) has been often detected in the
range $\approx$ 0.70--0.80 keV;  ($iii$) the observed blueshift of the
absorption lines imply they are originated in an outflow with  mean
velocities in the range $\approx$ 100--2500 km/s.

However, two important issues regarding the warm absorbers still
remain open, i.e.:  the exact location of the absorbing medium
and the physical connection between the absorbing materials in the UV
and X--ray band.  A likely range of distances of the ionized absorbing
medium from the central source is $r$ $\approx$ 0.01--10 pc (Krolik
2002; Behar et al. 2003).

By correlating {\it HST} and {\it ASCA} data Crenshaw
et al. (1999) found that all Seyfert galaxies  with evidence of UV
absorption also exhibit X--ray ionized absorption features.
Furthermore, some sources in which UV and X--ray absorbing components
share the same outflow velocity space have been observed (Blustin et
al. 2003; Collinge et al. 2001; Kaastra et al. 2002). However, no
clearcut identification of an X--ray warm absorber with an UV
counterpart has been reported yet (see Crenshaw, Kraemer \& George
2003 for a review).
  
Unfortunately, the sensitivity of current grating spectrometers allows
us to analyze just a handful of very bright  X--ray sources. Therefore
CCD cameras spectroscopy still
 represents the best available
tool  to deduce information about the physical properties of the gas
in the immediate surroundings  of the central X--ray emitting region
for most AGNs (i.e. those with a \fhx~\simlt~1 mCrab\footnote{1 Crab 
corresponds to a 2--10 keV flux of $\sim$ 2 $\times$ 10$^{-8}$
\cgs.}).  We present here the analysis of an \epic~observation of
Markarian 304 (\fhx~$\approx$ 0.1 mCrab).\\

Markarian 304 (MKN~304, PG2214$+$139) is a Seyfert galaxy at $z$ =
0.066.  Due to the presence of broad emission lines in its optical
spectrum, MKN~304  has been classified as a Type 1 object (Osterbrock
1977). $IUE$ spectroscopy of this source revealed the presence of
strong bump in the UV (Clavel \& Joly 1984), without absorption lines. 
On the contrary, a recent $FUSE$
observation of MKN~304 (Kriss 2002) disclosed broad absorbing features
(with a velocity spread of 1500 km/s) due to multiple OVI absorption
components blended together.

%X--ray history of MKN~304.
MKN~304 was discovered as X--ray emitting source by {\it Uhuru}
(Tananbaum et al. 1978). Later X--ray observations of this source were
performed by {\it Einstein} (Kriss, Canizares \& Ricker 1980) and {\it
ROSAT} (Rachen et al. 1996): both  reported evidences of heavy
obscuration, with a very flat slope ($\Gamma \sim$ 0.07) in the soft
X--ray band.  Surprisingly, MKN~304 was not targeted by {\it ASCA}
and {\it BeppoSAX} and, therefore, no further study of 
its unusual flat continuum was performed.

Throughout this paper we assume a flat {\em $\Lambda$CDM} cosmology
with ($\Omega_{\rm M}$,$\Omega_{\rm \Lambda}$) = (0.3,0.7) and a
Hubble constant of 70 km s$^{-1}$ Mpc$^{-1}$ (Bennett et al. 2003).

\section{\xmm~observations and data reduction}
MKN~304 was observed by the European Photon Imaging Camera ({\it
EPIC}) on board the \xmm~space observatory (Jansen et al. 2001)
between 2002 May 12 15:46:29 (UT) and 2002 May 13 00:56:34 (UT).  The
observation was performed with the \epic~\pn~camera (Struder et
al. 2001) operating in  Full Window Extended mode and with the
\epic~\mos~cameras (Turner et al. 2001)  operating in Full Window
mode.  All the three \epic~cameras were equipped with the ``Thin''
blocking optical filters.

The data files  were reduced and analyzed using the standard Science
 Analysis System ($SAS$) v.5.4.1. We used the  EPCHAIN and  EMCHAIN
 tasks for processing of the raw \pn~and \mos~data files,
 respectively, in order to generate the relative linearized event
 files.  X--ray events corresponding to patterns 0--12 (i.e. single--,
 double--, triple--, and quadruple--pixel events) for the \mos~cameras
 were selected; for the \pn, patterns 0--4 events (i.e. single-- and
 double--pixel events) were used. The subsequent event selection was
 carried out taking into account  the most updated calibration files
 at the time the reduction was performed (September 2003). All known flickering and
 bad pixels were removed.  The event lists were furthermore filtered to
 ignore periods of high  background flaring (see Appendix A).  After
 this screening process (and taking into account satellite
 dead  times) the effective net exposure times for the {\em MOS1},
 {\em MOS2} and \pn~detectors are 33.4, 35.5 and 29 ks, respectively.
 No exposure is affected by pile--up.  The peak of the X--ray emission
 is located at $\alpha_{2000}$ = 22$^{h}$17$^{m}$12.1$^{s}$ and
 $\delta_{2000}$ = $+$14$^{\circ}$14$^{\prime}$21.7$^{\prime\prime}$,
 i.e. at a distance of 1.5  arcsecs from the coordinates of the
 optical nucleus of MKN~304 (Clements 1981).
This is consistent with typical attitude reconstruction uncertainties.

Since the cleaned {\em RGS} data did not yield enough photons
for  a meaningful spectral analysis, they are not discussed
in the following.

The source spectra were extracted from the final filtered event lists
using a circular region of 24 and 21.5  arcsecs radius for \pn~and
\mos, respectively, centered on the observed position of MKN~304.
Backgrounds were estimated from source--free similar regions close to
the source on the same CCD.  Response matrices and ancillary response
files were generated using the RMFGEN and ARFGEN tools in the $SAS$
software.  Since difference between the {\em
MOS1} and {\em MOS2} response matrices are  a few percent, 
we created a combined \mos~spectrum and
response matrix.  The \pn~and
\mos~spectra were then fitted simultaneously.  Both spectra were
grouped to give a minimum of 40 counts per bin in order to apply
$\chi^{2}$ statistics.  Given the current calibration uncertainties
and the detector sensitivities, events outside the 0.3--12 keV range
were ignored in the \pn~spectrum, while, for the \mos, we retained
the 0.6--10 keV band.
\section{X--ray spectral analysis}

Spectral analysis was performed using XSPEC v.11.2 (Arnaud 1996).  All
the models presented in this Section and in the following one include
absorption due to the line--of--sight Galactic column of \nh~= 4.96
$\times$ 10$^{20}$ \cm2~(Dickey \& Lockman 1990).  The photoelectric
absorption cross sections were taken from Morrison \& McCammon (1983).
The quoted errors refer to the 90\% confidence level for one
interesting parameter (i.e. $\Delta\chi^2$ = 2.71; Avni 1976).

The 0.5--2 keV and 2--10 keV \pn~lightcurves were visually inspected
for checking variability.  We did not detect any significant change in
both count rates during the $\approx$ 9 hr observation.

\begin{figure}
\begin{center}
%\leavevmode
\psfig{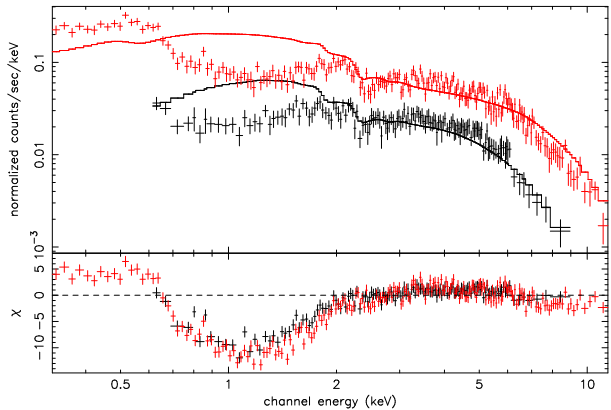}
\caption{{\it Upper panel}: Continuum power law fit to the 2--12 keV
 band of the \pn~(top) and \mos~(bottom) spectra of MKN~304,
 extrapolated over 0.3--12 (0.6--12) keV for \pn(\mos). {\it Bottom
 panel}: Residuals in units of standard deviations.}  
\label{fig:ext2soft}
\end{center}
\end{figure}

We first fitted the \epic~spectra with a power law model in the 2--12
keV band.  This simple model  yielded a very poor fit (\xnu =
469(256)) and  an extremely flat photon index of $\Gamma \approx$
0.95. Indeed the convex shape of the data--to--model residuals
suggests the  presence of heavy obscuration.  Fig.~\ref{fig:ext2soft}
displays the results of this fit extrapolated over the 0.3--12 keV
bandpass.  The large deficit in the 0.5--2 keV band clearly indicates
the presence of a strong absorption, while the systematic positive
residuals at \simlt~0.6 keV may reveal an emission component in excess
of the underlying absorbed continuum.  Fitting the data over the
0.3--12 keV band with a power law yielded a \xnu~= 2483(367) (see
model A in Table~\ref{tab:results}).

We thus accounted for this ionized absorber component with the
multiplicative {\bf absori} model  in XSPEC (Done et al. 1992). In
this model the state of the warm  absorber is a function of the
ionization parameter $\xi$ defined as $\xi$ = $L$/$n$$r^{2}$, where
$L$ is the isotropic luminosity of the ionizing source in the interval
5 eV to 300 keV, $n$ is the number density of the warm plasma  and $r$
is the distance between this latter and the central source. In the fit
solar elemental abundances were assumed and  $\xi$ was left as a free
parameter. On the contrary, we fixed the temperature of the absorber
to T = 1.5 $\times$ 10$^{5}$ K  since the statistics did not allow
to simultaneously and effectively constrain $T$ and $\xi$. Such a
value of $T$ falls in the typical range observed for a warm absorber
(Reynolds \& Fabian 1995; Krolik 2002 and reference therein).  We also
performed a further set of fits assuming values of temperature in the
range 5 $\times$ 10$^{4}<T<$ 10$^{6}$ K, from which turned out that
the ionization parameter  remains almost constant (i.e. $\xi \approx$
90--100 \cgs) to the variations of temperature. This fit (indicated
as model B in Table~\ref{tab:results}) resulted in a dramatic
statistical improvement i.e. at $>$ 99.99\% confidence level
according to an $F$--test once compared to model A,  with a \xnu~=
617(365).  Nevertheless, the value of $\chi^2$/$\nu$ $\approx$ 1.7
remains unacceptable and a large absorption feature  is evident in the
soft portion of the spectrum (see Fig.~2).  The breadth and depth of this feature
suggest the existence of another  ionized absorption component with a
different value of $\xi$. Furthermore the derived slope $\Gamma$ =
1.40$\pm$0.04 is  significantly flatter than usually observed in
unabsorbed Seyfert galaxies (Malizia et al. 2003) and quasars
(Piconcelli et al. 2003), i.e. $\langle\Gamma\rangle$ $\approx$ 1.9.

We then added another {\bf absori} component in the spectral model.
First we fitted the spectrum fixing the temperature of this new
component to the same value of the previous one, i.e. T = 1.5 $\times$
10$^{5}$ K. The fit gave a minimum \xnu~= 477(363), with a statistical
improvement significant at $>$ 99.99 confidence level. However, an
even better $\chi^2$ value ($\chi^2$ = 441) was obtained  fixing the
temperature of the second ionized component to a lower value, in
particular  with T = 3 $\times$ 10$^{4}$ K.  The resulting spectral
parameters of this fit (Model C) are also listed in
Table~\ref{tab:results}.  The $\chi^2$ value was reduced by
$\Delta\chi^2$ = 176 respect to Model B, thus the addition of the
low-temperature ionized absorption term is significant at a level
larger then 99.99\% confidence level  ($F$--test value = 72).
%%%%%%%%%%%%%%%%%%%%%%%%%%%%%%%%%%%%%%%%%%%%%%%%%%%%%%%%%%%%%%%%%%%%%%%%%%%%%%%
\begin{figure}
\begin{center}
\psfig{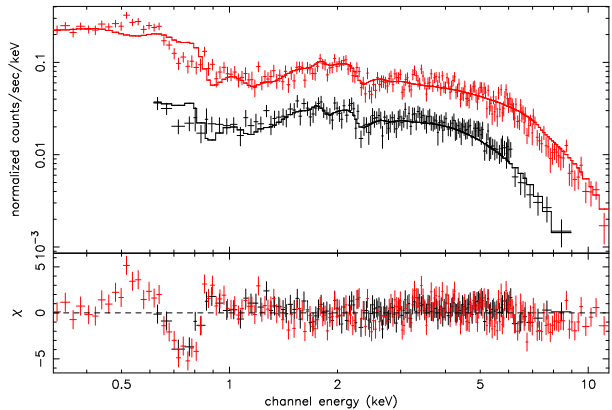}
\caption{{\it Upper panel}: The \pn~(top) and \mos~(bottom) spectra of
MKN~304 fitted by a single warm absorber model (model B in Table~1).
{\it Bottom panel}: Residuals in units of standard deviations. They
clearly indicate that this model cannot account for the complex
spectrum of MKN~304.}
\label{fig:modelB}
\end{center}
\end{figure}
%%%%%%%%%%%%%%%%%%%%%%%%%%%%%%%%%%%%%%%%%%%%%%%%%%%%%%%%%%%%%%%%%%%%%

However the reported value of $\chi^2$/$\nu$ = 1.22 (null hypothesis
probability 0.003) is not yet very satisfactory and positive
systematic residuals are still present in the data--to--model ratio
below 1 keV.
%%%%%%%%%%%%%%%%%%%%%%%%%%%%%%%%%%%%%%%%%%%%%%%%%%%%%%%%%%%%%%%%%%%%%%%%%%%%%%%
\begin{figure}
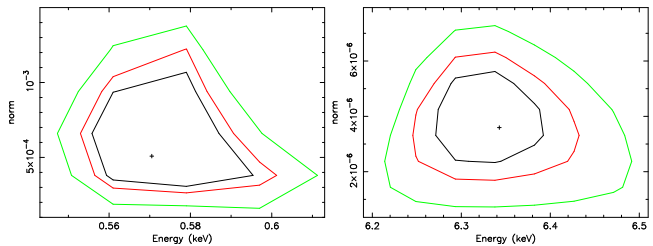

\begin{center}
\centerline{
\psfig{figure=epico_f3.ps,width=42.truemm,angle=-90}
\psfig{figure=epico_f4.ps,width=42.truemm,angle=-90}}
\caption{{\it Left panel}: contours at the 68\%, 90\% and 99\% confidence level for energy and
flux of the He--like oxygen line.  
{\it Right panel}: contours at the 68\%, 90\% and 99\% confidence level for energy and
flux of the iron line. Both fluxes are in units of photons$/$cm$^{2}$$/$s.}
\end{center}
\label{fig:2lines}
\end{figure}
%%%%%%%%%%%%%%%%%%%%%%%%%%%%%%%%%%%%%%%%%%%%%%%%%%%%%%%%%%%%%%%%%%%%%

%%%%%%%%%%%%%%%%%%%%%%%%%%%%%%%%%%%%%%%%%%
%%%%%%%%%%%%%%%%%%%%%%%%%
%%%%%%%%%%%%%%%%%%%%%%%%%%%%%%%%%%%%%%%%%% $^{+}_{-}$
\begin{table*}
\baselineskip=25pt
\begin{center}
\begin{tabular}{ccccccccc}
\hline
\multicolumn{1}{c} {Model}& \multicolumn{1}{c}
{$\Gamma$}& \multicolumn{1}{c} {\nhh}& \multicolumn{1}{c}
{$\xi^{T5}$}& \multicolumn{1}{c} {\nhc}& \multicolumn{1}{c}
{$\xi^{T4}$}& \multicolumn{1}{c} {Line Energy}& \multicolumn{1}{c}
{$\rm{\chi^{2}(d.o.f.)}$}\\

&
&($10^{22}$~cm$^{-2}$)&($\rm{erg~cm^{-2}~s^{-1}}$)&($10^{22}$~cm$^{-2}$)&($\rm{erg~cm^{-2}~s^{-1}}$)&(keV)
&  \\ \hline A &0.58$^{+0.02}_{-0.02}$&--&--&--&--&--&2483(367)\\ B
&1.40$^{+0.04}_{-0.04}$&6.4$^{+0.3}_{-0.3}$&92.9$^{+9.8}_{-9.7}$
&--&--&--&617(365)\\ C &1.64$^{+0.06}_{-0.06}$&8.2$^{+0.7}_{-0.7}$&110.8$^{+8.4}_{-7.7}$&0.55$^{+0.08}_{-0.08}$
&3.8$^{+2.6}_{-1.4}$&--&441(363)\\ D &1.92$^{+0.08}_{-0.08}$&
10.9$^{+0.9}_{-1.0}$&112.7$^{+29.0}_{-17.5}$&1.3$^{+0.3}_{-0.3}$&2.9$^{+1.2}_{-0.9}$&--&372(362)\\
E
&1.88$^{+0.04}_{-0.04}$&8.9$^{+0.5}_{-0.5}$&89.3$^{+13.9}_{-12.0}$&1.7$^{+0.2}_{-0.4}$&5.9$^{+2.4}_{-0.9}$&6.34$^{+0.06}_{-0.06}$/0.57$^{+0.02}_{-0.01}$&335(358)\\
\hline
\end{tabular}
\end{center}
\caption{X--ray spectral model fitting results. Parameters marked with
T5(T4) are referred to the high--(low--) temperature component of the
warm absorber. In model E the photon index of the unabsorbed power law
was fixed to the value of the primary absorbed continuum. See text for
details.}
\label{tab:results}
\end{table*}
%%%%%%%%%%%%%%%%%%%%%%%%%%%%%%%%%%%%%%%%%%%
%%%%%%%%%%%%%%%%%%%%%%%%%%%%%%%%%%%%%%%%%%%
%%%%%%%%%%%%%%%%%%%%%%%%%%%%%%%%%%%%%%%%%%%

We added a second power law accounting for this additional soft X--ray
emitting component.  The relative photon index was constrained to be
that of the primary continuum but the normalization was left free to
vary.

We obtained again a very significant improvement in the fit quality at
the $>$ 99.99\% confidence level ($F$--test = 66) with reduced
$\chi^2$ = 373 for 362 degrees of freedom (see model D in Table 1).
The resulting photon index is $\Gamma$ = 1.92$\pm$0.08 and the
normalization of the  unabsorbed power law  is $\approx$ 4\% of the
absorbed power law component. The best fit column densities of the two
ionized absorbers increased to \nhc~= 1.3$\pm$0.3 $\times$ 10$^{22}$
\cm2~and   \nhh~= 10.9$^{+0.9}_{-1.0}$ $\times$ 10$^{22}$ \cm2~for
the lukewarm (i.e. with T = 3 $\times$  10$^{4}$ K) and hot (i.e. with
T = 1.5 $\times$ 10$^{5}$ K) component, respectively.  The ionization
parameters turn out to be $\xi^{T4}$ = 3.8$^{+2.6}_{-1.4}$ \cgs~and
$\xi^{T5}$ = 110.8$^{+8.4}_{-7.7}$ \cgs.
  
We attempted an alternative fit modelling the soft excess with a blackbody component 
attenuated by both ionized absorbers, and with all fit parameters free to vary, 
which yielded a minimum \xnu~= 399(361) and, therefore, significantly worse than model D.
We also tried to substitute the ``lukewarm'' ionized absorbing component in model D with a neutral
one, but the quality of the fit resulted very poor  with a final \xnu~= 532(363)
(null hypothesis probability 1.5 $\times$ 10$^{-8}$).

Model D provides an excellent and physically plausible parametrization
of the overall \epic~spectrum of MKN~304,  with a null hypothesis
probability equal to $\approx$ 0.5. Visual inspection of the
data--to--model ratio residuals suggests the presence of two emission
features, respectively, at $\sim$ 6 keV and $\sim$ 0.5 keV
(observer--frame).  The former is likely due to the fluorescence
emission from the iron K--shell as commonly observed in most of
Seyfert galaxies (Nandra et al. 1997).   The latter is likely due to
the helium--like oxygen triplet of emission lines centered at 0.57 keV
(rest--frame) as found in other Seyfert galaxies (e.g. Kaastra et
al. 2002).  Thus the data led us to add two narrow gaussian lines in
model D to account for them.  The best fit rest-frame energy of the Fe
K$\alpha$ line is 6.34$\pm$0.06 keV with an equivalent  width of EW =
79$\pm$35 eV.  This line is significant at 99.9\% confidence level
($\Delta\chi^2$ = 13 for two additional parameters; $F$--test  value =
7.3), the centroid energy corresponds to low ionization states,
i.e. FeI--XV (Makishima 1986). We obtained an upper limit for the line
width of $\sigma_{K\alpha} <$ 0.18 keV.  The rest--frame position of
the soft X--ray line ($E$ = 0.57$^{+0.02}_{-0.01}$ keV) is fully
consistent with OVII  emission. Given the reduction of $\chi^2$ of
$\Delta\chi^2$ = 25 for two additional degrees of freedom,  the
detection of this line is significant at $>$ 99.99\% confidence level
based on an $F$--test. In Fig.~\ref{fig:2lines} are shown the 68.3\%,
90\% and 99\% confidence levels for two interesting parameters between the
line energy and the flux of both emission features assuming model E in Table~1.
 
\begin{figure}
\begin{center}
\psfig{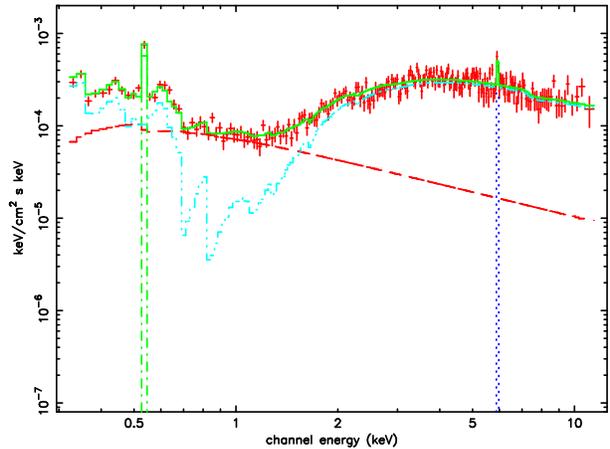}
\caption{Unfolded best--fitting \epic~spectrum of MKN~304 fitted with
two power laws plus a double warm absorber component and two gaussian
lines (model E in Table~1).  The photon index of the unabsorbed power
law is fixed to the value of the primary absorbed continuum  (see text
for details).}
\label{fig:thebest}
\end{center}
\end{figure}

Fig.~\ref{fig:thebest} shows the \pn~and \mos~spectra and the spectral
fit which includes the two gaussian emission  line. We consider this
spectral model (Model E in Table~1)as the best fit to the \epic~data
of MKN~304.  This fit yields a minimum \xnu~= 335(358) with a null
hypothesis probability of 0.82.
%%%%%%%%%%%%%%%%%%%%%%%%%%%%%%%%%%%%%%%%%%%%%%%%%%%%%%%%%%%%%
%%%%%%%%%%%%%%%%%%%%%%%%%%%%%%%%%%%%%%%%%%%%%%%%%%%%%%%%%%%%%%%%%%%%%

Once Model E was assumed, we measured a flux of \fsx~= 3.06 $\times$
10$^{-13}$ \cgs~and \fhx~= 3.26 $\times$ 10$^{-12}$ \cgs~ in the low
(0.5--2 keV) and high (2--10 keV) energy bands, respectively.  After
correction for both Galactic absorption and warm absorber column
densities, these correspond to luminosities of \lums~= 3.9  $\times$
10$^{43}$ \ergs~and \lum~= 4.8 $\times$ 10$^{43}$ \ergs, respectively.
During the \xmm~observation the soft X-ray flux of MKN~304 was therefore
a factor of 0.5 lower than measured\footnote{In order to compare the fluxes measured
by different satellites we assumed a power law model with $\Gamma$ = 1.8 and \nh = \nhgal.}
by {\it Einstein} on 1979 and a factor 
of 1.4 higher than the value obtained by {\it ROSAT} on 1992.
Unfortunately, due to the lack of spectral information we cannot
investigate the possibility that these variations are linked to changes
in the column density of the observer or, alternatively, caused by the
activity cycle of the X--ray continuum source.
%%%%%%%%%%%%%%%%%%%%%%%%%%%%%%%%%%%%%%%%%%
%%%%%%%%%%%%%%%%%%%%%%%%%
%%%%%%%%%%%%%%%%%%%%%%%%%%%%%%%%%%%%%%%%%%
%%%%%%%%%%%%%%%%%%%%%%%%%%%%%%%%%%%%%%%%%%%
%%%%%%%%%%%%%%%%%%%%%%%%%%%%%%%%%%%%%%%%%%%

\section{Discussion}
The spectrum of MKN~304 unveiled by the \xmm~observation discussed
 in this paper is complex both in the soft  and in the hard energy 
band (Fig.~1).

Our analysis fully confirms previous results on narrower energy ranges
which reported a very flat continuum for this source, and  suggested
the existence of heavy obscuration. However, due to the limited
bandpass and the poor sensitivity of the previous observations, 
it was impossible to accurately investigate the properties
of the absorption, and discriminate between a cold (neutral)
or warm (ionized) medium.    This \xmm~observation of
MKN~304 has clearly revealed that the strong obscuration
(\nh~$\approx$ 10$^{23}$ \cm2) occurs in an ionized gas.

We have indeed found that the best fit model to the \epic~data
consists of a two--phase warm absorber with a high-- and a
low--ionized component, a faint soft excess component and two emission
lines at $\sim$ 0.57 keV and $\sim$ 6.4 keV,  respectively (see Model
E in Table~1).

\subsection{Warm absorber} 
\label{sect:warm}

We fixed {\it a priori} the temperature values at T = 1.5 $\times$
10$^{5}$ K and T = 3.0 $\times$ 10$^{4}$ K for the two ``phases'' of
the ionized plasma since they represent typical values observed in
sources with a multi--phase warm absorber (Netzer et al. 2003; Kaastra
et al. 2002).  From the spectral analysis we derived $\xi^{T5}$ =
89.3$^{+13.9}_{-12.0}$ \cgs~and $\xi^{T4}$ = 5.9$^{+2.4}_{-0.9}$
\cgs~for the hot and the cold component, respectively (see model E in
Table 1).

Our findings for MKN~304 appear to be in agreement with  a
``multi--zone'' structure for the ionized absorber in Seyfert galaxies
as discovered by high resolution spectroscopic observations recently
performed  with grating spectrometers on--board of \xmm~and \chandra.
They indeed resolved the complex structure of the AGN circumnuclear
absorbing material into a multi--phase plasma with  several  orders of
magnitude spread in $\xi$  and systematic blueshifts (Blustin et
al. 2003; Kaastra et al. 2002).   This observational evidence leads to
interpret the warm absorber phenomenon essentially in terms of a
photoionized  outflowing wind consisting of different inhomogeneous
co--existing regions (or shells) characterized by a broad distribution
in $\xi$, temperature and density (Krolik 2002; Behar et al. 2003;
Elvis 2000).

 Interestingly, while our best--fit values for ionization parameters
of the two absorbing components are similar to those found in other
Seyfert 1 galaxies (Kaastra et al. 2002; Blustin et al. 2003),  the
corresponding column densities (\nhh~$\approx$ 9 $\times$ 10$^{22}$
\cm2~and \nhc~$\approx$ 2 $\times$ 10$^{22}$ \cm2) are amongst the
highest  seen by \chandra~and \xmm~so far.

Given these high values of \nh, photoionization edges are expected to
dominate the absorption features in MKN~304.  In fact, at high column
densities the UTA absorption lines saturate. Edges (alongside with
$2p-nd$ (with $n >3$) absorption lines)  provide the major
contribution  to the absorption spectrum (Behar, Sako \& Kahn 2001).
Furthermore as shown in Fig.~\ref{fig:ext2soft}, the absorption
feature in MKN~304 is remarkably broad and deep and, hence, hard to
reconcile with what simply expected  from UTA\footnote{Nonetheless, it
is also worth noting that the colder ionized component has values  of
\nh, $\xi$ and $T$ approximately consistent with the production of UTA
features around 0.75 keV. Unfortunately, the lack of an adequate
energy resolution does not allow us to further test this hypothesis.}.

Besides absorption, also emission lines are expected to arise from the
warm medium.  Due to the underlying illuminating bright continuum and
the forest of absorption lines such emission features are however
difficult to detect in low--resolution data. In particular He--like
triplets of O  and Ne are usually expected.  The OVII line detected at
0.57 keV in the spectrum of MKN~304 is therefore consistent with
reprocessing in the warm gas.

However, in sources showing both absorption and emission lines the
former usually appear to have different  velocities respect to the
latter (and in many cases consistent with the galaxy rest--frame,
e.g. Collinge et al. 2001), so it is not clear if emitter and absorber
systems are intrinsically the same.  For instance, Kaastra et
al. (2002) found the optical Narrow Line Region as the most likely
origin of OVII and Ne IX emission  lines observed in
NGC~5548. Unfortunately, \epic~data do not have enough resolution to
shed light on this issue.

We used CLOUDY (Ferland 2001) to further test the idea of a possible
origin of the oxygen emission line from the same {\it warm} medium
responsible of the absorption in MKN~304. We employed the table
``AGN''  ionizing continuum, with an X--ray photon index fixed to
1.88, a hydrogen column density set to 9 $\times$ 10$^{22}$ \cm2, and
an ionization parameter equivalent to the highest one obtained  from
XSPEC. For this high ionization state of the gas, the transmitted
X-ray spectrum is only marginally dependent on the gas density, the
EUV continuum and the UV/X-ray flux ratio. Following the results of
the XSPEC fits, the total output spectrum from CLOUDY has been
computed as the sum of the incident and transmitted spectra with
relative contributions of 5 and 95\%, respectively. In this composite
spectrum the most prominent emission feature is due to OVII, whose EW
($\sim$ 110 eV) is in excellent agreement with that inferred by our
fit, i.e. EW = 105$\pm$18 eV. Other emission features in the model
(i.e. NeIX, NVI, MgXI) are several times weaker, except for the  Fe
K$\alpha$ line, whose EW is similar to that of OVII (see Sect. 4.3.).
Therefore, we conclude that although this result does not ultimately
proof the origin of the He--like oxygen line in the warm absorber, it
clearly supports such a scenario.

Finally, the combination of an unreddened continuum in the optical
band (H$\alpha/$H$\beta \sim$ 3; e.g. Osterbrock 1977) and a column
density of  \nh~$\sim$ 10$^{23}$ \cm2~suggests a dust--free X--ray
absorbing medium in MKN~304, maybe located  within the dust
sublimation radius (see Crenshaw, Kramer \& George 2003 and references
therein).

\subsection{Soft excess}

The soft excess component (whose normalization is $\approx$ 4\% of the
primary emission) detected in MKN~304  appears similar to that
observed in many X--ray obscured objects (Turner et al. 1997).  There
are two possibilities to explain the origin of such excess, i.e:
scattered$/$reflected emission from the photoionized gas  in the
warm absorber and ``partial--covering''.   We discuss in turn these
hypothesis in the following.

\xmm~and \chandra~results on Seyfert 2 galaxies (Sako et al. 2002 and
reference therein) suggest that the  soft X--ray emission in these
objects is the result of primary continuum scattering underlying a
blend of strong emission lines, mostly from H-- and He--ions of C, N,
O, Ne, and Fe.  Since warm absorber regions show physical properties
(i.e. $\xi$, column densities, temperatures)  similar to those
observed in the X--ray  photoionized emitting plasma in Seyfert 2
galaxies, it is likely that they are different manifestations of the
same phenomenon (Kinkhabwala et al. 2002) simply seen at different
inclination angles.  Within this scenario the soft excess found in
MKN~304 would be therefore  due to scattered$/$reflected
emission from the ionized outflowing plasma.

 Interestingly, Ogle et al. (2000), Bianchi et al. (2003) and Iwasawa
et al. (2003) detected indeed an extended soft X--ray emission
dominated by narrow emission lines overlapping the optical Narrow Line
Region in NGC~4151, NGC~5506 and NGC~4388,  respectively.  According
to this hypothesis, a significant contribution from this extended soft
X--ray component to the OVII emission line would be also expected (see
Sect.~4.1).

An alternative explanation of this ``soft excess'' could also be
some fraction of the primary continuum leaking  through the clumpy
obscuring material (i.e. along the line of sight the absorber does not
completely cover the nuclear source):  the  so--called
``partial-covering'' scenario. Furthermore, as mentioned in
Sect.~\ref{sect:warm}, clumpiness in the ionized absorbing gas appears
to be very likely.

Unfortunately, the present data do not allow to discriminate these
hypothesis.  The unabsorbed power law in model E (with a slope
identical to the primary continuum one) can be therefore considered
as a good approximation of the real (more complex) spectrum. However,
given the energy resolution of \epic~CCD cameras and the relatively
faintness of MKN~304, a  description of this soft X--ray component in
terms of many discrete emission features, as observed in the high
resolution spectra of some bright AGNs (see Kinkhabwala et al. 2002),
cannot be ruled out.

Finally we rule out the hypothesis of an origin of the soft excess
from reprocessing (i.e. reflection) by an ionized disk since this
interpretation is inconsistent with the observed centroid of the iron
line. In fact, assuming that the line arises from such an ionized disk
(as expected if a strong relection component is present; Ross, Fabian
\& Young 1999),  it should be emitted at energy higher than 6.4 keV
from iron at high ionization states.  On the contrary, we found an
energy of the  Fe K$\alpha$ emission line  strictly consistent only
with ``neutral iron'' (see below).

\subsection{Iron line}
We report the first clear evidence of an iron fluorescence emission
line around 6.4 keV in the spectrum of MKN~304.  The detection of a Fe
K$\alpha$ line at E = 6.34$\pm$0.06 keV constrains the ionization
parameter of the emitting medium to be $\xi$ \simlt~100 \cgs, higher
values of $\xi$ would indeed produce a feature peaked at higher
energies (Kallman \& McCray 1982) excluded by the present data.
Interestingly, both cold and hot warm absorber components that resulted in
our best fit (model E) to the \epic~spectrum have values of $\xi$
which match well with this constraint.  This suggests that a
significant component of the iron emission line may be originated in
the warm medium.

We have run the same CLOUDY model as in Sect.~4.1 to predict the
strength  of the Fe K$\alpha$ emission line.  The expected value
($\approx$ 110 eV) falls in the range we measured from the
\xmm~spectrum (i.e. EW = 79$\pm$35 eV).  So, interestingly, the iron
line detected in MKN~304 is consistent with an origin in the warm
absorber observed in this source.

We have also tried to fit the data with a model including a neutral
reflection Compton component ({\bf pexrav} model in XSPEC) in order to
test a possible origin of the iron line from the  optically--thick
accretion disk.  This spectral parametrization yielded a good fit with
a \xnu~= 335(357): however, there is no statistical improvement
respect to Model E.  Although the detection of cold reflection results
therefore only marginal, we cannot rule out its presence. The observed
spectral range (0.3--12 keV) of this \epic~observation is not adequate
to constrain the strength of this ``bump--like'' reprocessed emission
feature  peaked at $\approx$ 30--40 keV (Lightman \& White 1988).  The
resulting upper limit for the covering factor of the material
irradiated by the X--ray source is $R\equiv$$\Omega/2\pi <$ 0.7.  Both
the value of the equivalent width of the line and $R$ suggest that the
reflector -- if existing -- subtends a solid angle of  \simlt~$\pi$
steradian to the X--ray source, i.e. intercepts \simlt~50\% of the
primary continuum once a slab geometry for the reflecting medium is
assumed (George \& Fabian 1991).

Furthermore the narrow profile of the Fe K$\alpha$ line
($\sigma_{K\alpha} <$0.18 keV corresponding to $\Delta$$v$(FWHM)
\simlt~8000 km/s for the line velocity width) implies that even if the
reprocessing medium is the accretion disk, the fluorescence does not
take place  from its innermost (relativistic) regions.  A substantial
contribution to the Fe K$\alpha$ emission could be provided from the
putative molecular  torus located at 1--10 pc from the central X--ray
source as invoked in the Unification model of AGNs (Antonucci 1993).
In this case the material  is assumed to be Compton--thick (i.e. with
\nh~\simgt~1.5 $\times$  10$^{24}$  \cm2, see Matt, Guainazzi \&
Maiolino 2003), and the emission is due to the reflection of the
primary continuum off the inner walls of the torus into our
line--of--sight (Ghisellini et al. 1994).

The observed properties of the Fe K$\alpha$ line in MKN~304 appear to
be very similar to the results obtained by the spectral analysis of
the brightest Seyfert 1 galaxies using \xmm~(Pounds \& Reeves 2002)
and  \chandra~(Padmanabhan \& Yaqoob 2002) data.  These works indeed
shows the ``ubiquitous'' presence of a narrow iron line component
centered at 6.4 keV with an EW  $\approx$ 50--100 eV and
$\sigma_{K\alpha} <$120 eV in the X--ray spectra of the
low--luminosity (i.e. with \lum~$<$ 10$^{45}$ \ergs) AGNs.

\section{Summary}
The $\sim$30 ks \xmm~exposure presented here is the first  X--ray
broad--band (0.3--12 keV) observation of MKN~304. Our data analysis
has revealed that this Seyfert 1 galaxy has a complex spectrum,
dominated by heavy ionized obscuration. 
This warm absorber can be well fit by a ``two--zone'' gas with 
ionization parameter of $\xi$ $\approx$ 6~\cgs~and $\xi$ $\approx$ 90~\cgs, 
respectively. We estimate that the warm absorber has 
a column density of \nh~$\approx$ 9 $\times$ 10$^{22}$ \cm2~ in the 
high--ionization component and \nh~$\approx$ 2 $\times$ 10$^{22}$ \cm2~
in the low--ionization component. Such values are currently 
amongst the largest ones observed so far in Seyfert 1 galaxies by \xmm~and \chandra.

A particularly interesting result of our analysis is the detection
of an emission feature at $\approx$ 0.57 keV, likely due to the 
helium--like oxygen from the warm absorber itself or 
regions much farther out from the central BH (i.e. the optical Narrow 
Line Region).  

We also report the first detection (significant at 99.9\%) of an
iron line in this source. The combination of the energy centroid ($\approx$ 
6.4 keV) and the upper limit of the line width ($\sigma_{K\alpha} <$ 0.18 keV) 
suggests an origin from ``cold'' (i.e. FeI--XV) matter distant from 
the inner disk region. Alternatively, the line strength is 
consistent with the emission arising from the warm absorber itself.

Finally, we have also found a weak (4\% of the primary continuum) soft
excess.  The presence of this excess could be
explained by either emission/scattering from a warm gas or partial covering,
or a combination of them.

\section*{Acknowledgments}
We thank the anonymous referee for careful reading and helpful comments.
We are grateful to all the members of the \xmm~team for their great efforts
in performing the observations and supporting the data analysis.
This paper is based on observations obtained with \xmm, an ESA science
mission with instruments and contributions  directly funded by ESA
Member States and the USA (NASA).  This research has made use of the
NASA$/$IPAC Extragalactic Database (NED) which is operated by the Jet
Propulsion Laboratory,  California Institute of Technology, under
contract with the National Aeronautics and Space Administration.

When our paper was in an advanced stage of
the refereeing process we became aware of the publication  of a  paper
by Brinkmann et al. (2004) which discusses the same observation.
Their results are broadly consistent with ours insofar as they found
strong evidence   for ionized material along the line of sight.
However, the application of the high background filtering procedure
described in the Appendix allowed us to provide a more detailed
description of the emission and absorption features in the
spectrum. In particular, this leads us to unveil significant evidence
for a multizone warm absorber in MKN 304

\appendix

\section[]{High background filtering procedure} 
\xmm~experiences high ``flaring'' background  periods
due to the crossing  of clouds of high--energy particles (mainly soft protons),
which are focussed through the mirrors to the detectors (Lumb et al. 2002). 
Time intervals affected by background flares are
essentially useless for the scientific analysis. 
  The standard method  to filter out  these high
background  periods consists  in rejecting  periods of high count rate (CR)
at energies \simgt~10~keV (see \xmm~Science Analysis System Users Guide v2.1;
Loiseau 2003).  The recommended thresholds  for the
background count rate  are 0.35~count/s and 1~count/s for  the MOS and
the PN camera, respectively.   
This approach (i.e. using
fixed  threshold  for  the CR of the background) well adapts
for the 
detection of faint sources but it could produce very
conservative results in the case of high--flux sources. 

\begin{table}\baselineskip=25pt
\begin{center}
\begin{tabular}{ccc}
\hline
\multicolumn{1}{c} {}&
\multicolumn{1}{c} {Standard method}&
\multicolumn{1}{c} {MaxSNR method}\\
\hline
Final exposure&6.7 ks&29 ks\\
Obs. countrate&0.50 $\pm$0.01&0.480 $\pm$0.005\\
$SNR$         &53    &103\\
Total counts  &2971&16077\\
(0.3--12 keV) &   &\\
\hline
\end{tabular}
\end{center}
\caption{Comparison between the results obtained applying
the ``standard'' method and the ``MaxSNR'' method presented in this paper 
for the high background filtering
in MKN~304. Data are reported for the \pn~camera.}
\label{tab:2}
\end{table}
%%%%%%%%%%%%%%%%%%%%%%%%%%%%%%%%%%%%%%%%%%%%%%%%%%%%%%%%%%%%%
\begin{figure}
\begin{center}
\psfig{figure=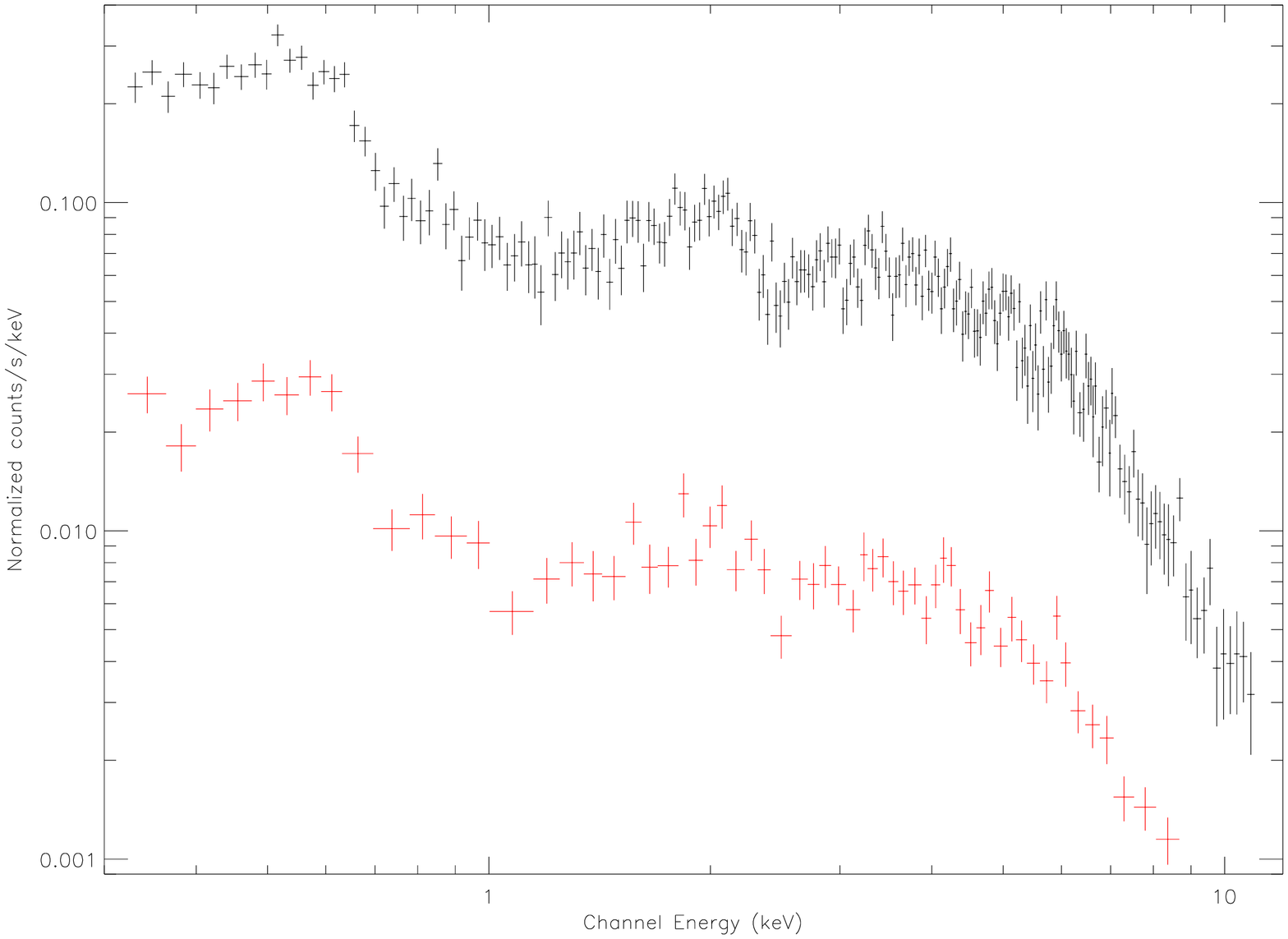,width=80truemm,angle=90}
\caption{Comparison between the \pn~spectrum of MKN~304 obtained
using the MaxSNR filtering method ({\it top}) and the standard filtering
procedure ({\it bottom}). After the filtering the exposure times result
of 29 ks and 6.7 ks, respectively.} 
\label{fig:snr}
\end{center}
\end{figure}
%%%%%%%%%%%%%%%%%%%%%%%%%%%%%%%%%%%%%%%%%%%%%%%%%%%%%%%%%%%%%%%%%%%%%
In the reduction of the MKN~304 data, we thus apply the following
alternative  procedure (``MaxSNR'' hereafter) to reject  the high background level
periods. The underlying idea basically consists of  filtering out
just those time  intervals for  which  the count  rate of  the
background reach values so high that the SNR of the source
does not significantly improve (or even worses) by including such
time intervals in  the analysis.  
To this aim, lightcurves of the source and
background region, i.e. a proper annulus around the source large enough to 
provide a good estimation of the background (with a 120 arcsec radius in the case of MKN~304), 
are extracted in the energy band  
for which the
SNR should to be maximized.  In the  analysis of  MKN~304 data, we have
taken into account  the 0.5--1.3~keV (0.6--1.3 keV)  band for \pn(\mos)
in order to optimize the SNR in the band where most of the prominent 
warm absorber features are located.

On the contrary, the ``standard'' filtering method 
requires to extract a lightcurve of the entire  
field of view for energies above $E >$ 10 keV.

Lightcurve  datapoints for the  source  and  background are
sorted for  increasing values of  the CR of  the
background region.  The  cumulative  signal--to--noise (SNR$_{cum}$)
distribution function is then calculated.

The SNR$_{cum}$ as a function  of the increasing background CR builds
a curve which monothonically rises up to a peak.  After  this  maximum
value,  the SNR$_{cum}$ distribution indeed flattens  or  even
decreases.  This value is the threshold  used to reject the high
background periods.  In fact, if included, these periods would not
provide any significant improvements to the final SNR of the spectrum
of the source, but it would increase the background contribution to
the raw source spectrum.

This  method  to filter  out  high  background  periods optimizes  the
duration of the observation useful for a meaningful scientific
analysis on the basis of the ratio between the count rates 
of source and background. It appears particularly appealing in the case
of bright source for which an even ``flaring'' background can 
represent just a negligible fraction of their CR.
It also has the advantage that the filtering is performed
in a particular energy band (chosen {\it a priori})
where the SNR need be maximized for the accurate study of peculiar 
spectral features.

Table~\ref{tab:2} lists the final values of exposure, total counts in
the 0.3--12 keV band, countrate and SNR for the $PN$ camera obtained
using the ``MaxSNR''  method compared to those derived from the
application of the ``standard'' method. In particular, the method
presented here yields a number of  counts which is about 5.5 times
larger than that resulted from the ``standard'' method.
Fig.~\ref{fig:snr} shows the comparison between the spectra obtained
using the two  filtering procedures: the higher quality of the
\pn~spectrum provided by the ``MaxSNR'' method is clearly evident.

\label{lastpage}
\bsp
\end{document}